\documentclass[amssymb,amsmath,aps,showpacs,twocolumn]{revtex4}
\usepackage{graphicx}
\usepackage{color}
\usepackage{soul}
\usepackage{latexsym}

\newcommand{\lsim}{\lesssim}
\newcommand{\gsim}{\gtrsim}

\def\lsim{\mathrel{\raise.3ex\hbox{$<$\kern-.75em\lower1ex\hbox{$\sim$}}}}
\def\gsim{\mathrel{\raise.3ex\hbox{$>$\kern-.75em\lower1ex\hbox{$\sim$}}}}

\def\beq{\begin{equation}}
\def\eeq{\end{equation}}
\def\beqn{\begin{eqnarray}}
\def\eeqn{\end{eqnarray}}
\def\bea{\begin{eqnarray}}
\def\eea{\end{eqnarray}}
\def\be{\begin{equation}}
\def\ee{\end{equation}}
\newcommand{\fslash}[1]{{#1 \kern -0.7em/ \kern 0.1em}}

\begin{document}

\voffset 1.25cm

\title{
Unique Higgs boson signature at colliders
}
\author{ Peng-fei Yin and Shou-hua Zhu}
\affiliation{Institute of Theoretical Physics, School of Physics,
Peking University, Beijing 100871, China}

\date{\today}

\begin{abstract}

The HyperCP collaboration has observed three events for the decay
$\Sigma^+ \rightarrow p \mu^+ \mu^-$. The three events may be
interpreted as a new narrow-width CP-odd scalar $a$ with the mass
$214.3 \pm 0.5$ MeV. Here $a$ decays dominantly into di-muon
($\mu^+\mu^-$). As the consequence of tiny mass difference between
$m_a$ and 2$m_\mu$ ($2 m_{\mu} \approx$ 211.3 MeV), di-muon will be
boosted to almost the same direction at colliders. Such kind of
di-muon events have been overlooked in the past experiments.
Provided that the precision data preferred light SM-like Higgs boson
$h$ decays dominantly into $a a$ other than into $b\bar b$, in order
to be consistent with null Higgs boson search at LEP, the
$h\rightarrow aa \rightarrow 4\mu$ ($2\mu^+ 2\mu^-$) will be the
unique Higgs boson signature which has not been noticed before. The
SM-like Higgs boson may hide itself from the usual analysis of LEP
and Tevatron experiments, which should be reanalyzed in the light of
new theoretical and experimental developments. In this paper, we
also investigate this unique Higgs boson signature at colliders and
conclude that the SM-like Higgs boson could be discovered with
rather low integrated luminosity, provided that the $h\rightarrow
4\mu$ reconstruction efficiency is not extremely low. It is not
impossible that such kind of unique Higgs boson $4\mu$ events are
now lurking in the existing LEP and/or Tevatron data.

\end{abstract}


\maketitle

\section{Introduction}
Understanding the mechanism of electro-weak symmetry breaking (EWSB)
is the primary goal for high energy experiments, namely
 Tevatron at Fermilab, the Large Electron-Positron (LEP),
the Large Hadron Collider (LHC) at CERN, and the proposed
International Linear Collider (ILC). In the standard model (SM) of
high energy physics, EWSB is realized via a weak-doublet fundamental
Higgs field. After EWSB spontaneously, namely Higgs field acquiring
a vacuum expectation value (VEV), only one neutral Higgs boson is
left in particle spectrum. The Higgs boson mass is theoretical
unknown within the SM. Therefore searching it in all mass regions is
necessary and great efforts have been put on it since the
establishment of the SM. The latest direct search at LEP sets the
lower bound of SM Higgs boson of $114.4$ GeV at 95\% confidence
level (CL) \cite{Barate:2003sz}. The Higgs boson can also affect
electro-weak observables through radiative corrections. Therefore
precise measurements of these observables can predict the Higgs
boson mass. Based on the global fit of data from LEP, SLD and
Tevatron, the Higgs boson mass in SM is predicted to be
$m_H=98^{+52}_{-36}$ GeV and $m_H <208$ GeV at 95\% CL using the top
quark mass $m_t=174.3\pm 3.4$ GeV \cite{Juste:2005dg}. However the
notorious three 3-$\sigma$ anomalies \cite{LEP-latest-talk} may
indicate new dynamics beyond the SM. Moreover excluding these
anomalies from the global fit data, the preferred even lighter Higgs
boson mass has shown certain tension with direct search limit at LEP
\cite{Chanowitz:2001bv}.

The constraint on Higgs boson mass from LEP direct search can be
greatly modified in the physics beyond the SM. For example, in the
next-to-minimal supersymmetric model (NMSSM) \cite{NMSSM} in which a
gauge singlet superfield is introduced, the SM-like CP-even Higgs
boson $h$ can mainly decay into light $a$ pair where $a$ is a
(mostly singlet) CP-odd Higgs boson \cite{Dermisek:2005ar}. Such
light $a$ may due to the approximate R-symmetry
\cite{Dobrescu:2000yn}. The relevant limit on $m_h$ can be deduced
from the measurements of more final states $Zh \rightarrow Z a a
\rightarrow Z \bar b b \bar b b$ or $Zh \rightarrow Z a a
\rightarrow Z \bar \tau \tau \bar \tau \tau$. The weaker limit of
$m_h$ can be obtained \cite{Dermisek:2005ar} primarily due to
dominance of $h \rightarrow a a$. Recently the authors of Ref.
\cite{Dermisek:2006wr} studied the specific scenario in which $m_h$
can be lighter than 100 GeV while $Br(h \rightarrow a a)>0.7$ and
$m_a < 2 m_b$.

The effects of such light CP-odd $a$ could be observed at low energy
experiments. Recently HyperCP Collaboration has observed three
events for the decay $\Sigma^+ \rightarrow p \mu^+ \mu^-$
\cite{Park:2005ek}. If the long-distance contributions are properly
included, it is possible to account for the branching ratio within
the SM \cite{He:2005yn}. However probability of all three events
with the same di-muon ($\mu^+\mu^-$) mass is less than one percent.
Thus it is natural to interpret three events from a new narrow-width
particle with mass $214.3 \pm 0.5$ MeV \cite{Park:2005ek}. The
theoretical investigations \cite{MWORKS} indicate that the new
particle can't be CP-even scalar or vector boson if they satisfy
also the constraints from $K^\pm \rightarrow \pi^\pm \mu^+\mu^-$,
$K_S \rightarrow \pi^0 \mu^+\mu^-$ and $B \rightarrow X_s
\mu^+\mu^-$. However the light CP-odd Higgs boson $a$ in the NMSSM
can be identified as the new light particle
\cite{He:2006fr,Hiller:2004ii}.

Assuming there exists the light CP-odd scalar with mass around 214
MeV as implied by HyperCP experiment, in this paper we will
investigate its phenomenological implications, especially for the
SM-like Higgs boson. Ref. \cite{Dermisek:2006wr} has investigated
the NMSSM scenario with O(GeV) $a$. For our purpose in section II,
we would like to study the NMSSM parameter space with the mass of
$a$ around 214 MeV. In section III, we will explore the unique
behavior of $a \rightarrow \mu^+\mu^-$ at colliders due to the tiny
mass difference between $m_a$ and $2 m_\mu$. In section IV, we carry
out the detail simulation of signals and backgrounds of SM-like
Higgs boson $h\rightarrow a a \rightarrow \mu^+\mu^-\mu^+\mu^-$ at
colliders. Section V is allocated to the conclusion and discussion.

\section{NMSSM parameter space with $m_a \sim 214$ MeV}

\begin{figure}[h]
\vspace*{-.07in}
\centerline{\includegraphics[width=3.0in,angle=0]{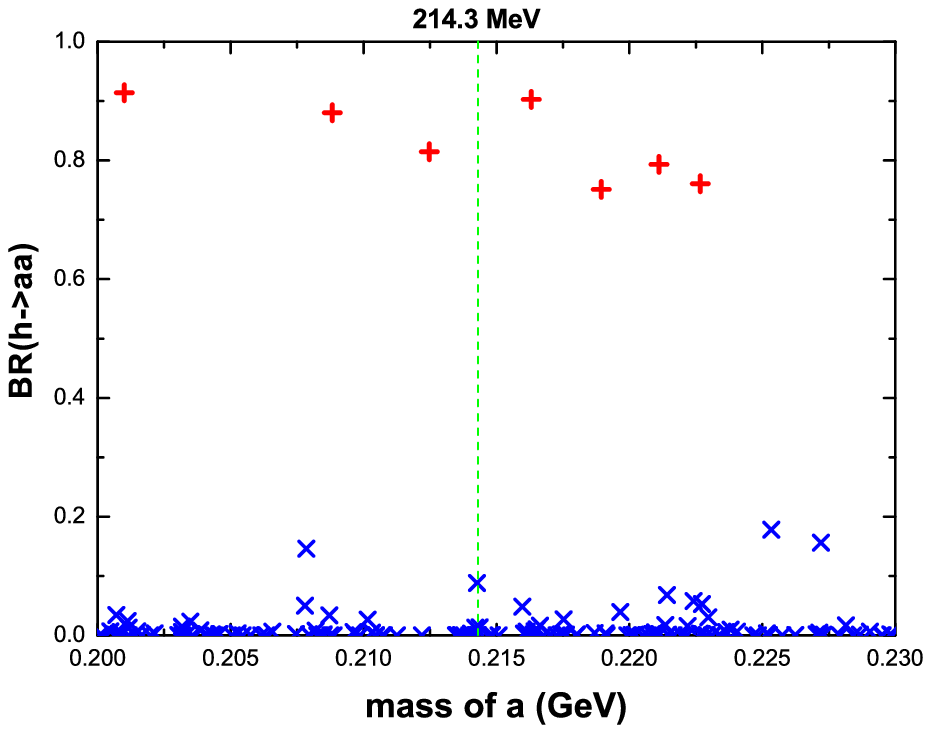}}
\vspace*{-.07in} \caption{BR($h\rightarrow aa$) as a function of
$m_a$ in the NMSSM with $\tan\beta=30\sim60$ , $|\mu|=100\sim300$
GeV , $A_t=1500 $ GeV, $M_{SUSY}=1000 $ GeV and
$M_{1,2,3}=100,300,500 $ GeV. The $+(\times)$ points indicate
$m_h<114 (>114)$ GeV.\vspace*{-.1in}} \label{NMHDECAY}
\end{figure}

We utilize NMHDECAY \cite{NMHDECAY:2006} to scan parameter space in
the NMSSM where $A_{\lambda}\approx A_{\kappa}\approx 0$ and ignore
fine-tuning constraints \cite{Dermisek:2005ar,Dermisek:2006wr}. We
choose large $\tan\beta$ and suitable $\mu$ as mentioned in Ref.
\cite{He:2006fr}. In Table~\ref{table1} we show three benchmark
points allowed by current experiment constraints embedded in
NMHDECAY, and all points have the $m_a$ around $214.3$ MeV. It is
clear that point 3 has large BR($h\rightarrow aa$) as we required.
Varying $\tan\beta$ and $\mu$ while keeping other parameters the
same with point 3,  we show the more allowed points in Fig.
\ref{NMHDECAY}. Both the table and figure show that there exist
possible parameter space as we required, i.e. BR($h\rightarrow aa$)
is large, the SM-like Higgs boson is around $O(100)$ GeV and CP-odd
$a$ is very light around $214.3$ MeV.

\begin{table}[htb]
\begin{tabular}{l|ccccccc}
\hline \hline
Points &\,BR($h\rightarrow aa$) &\, $m_a$  & $m_h$  &  $\lambda$  &  $\kappa$  &  $\tan\beta$  & \,$\mu$ \\
\hline
\,\,\,\,\,\,1 & \,$3.25\times10^{-5}$ &217.9 & 115.1 & 0.072 & 0.15 & 52.2 & 131.8 \\
\,\,\,\,\,\,2 & \,$8.82\times10^{-7}$ &214.4 & 115.3 & 0.026 & 0.05 & 33.6 & -151.1  \\
\,\,\,\,\,\,3 & 0.812 & 212.5& 88.4 & 0.067 & 0.024 & 33.7 & 130.0 \\
\hline \hline
Points & \,\,$M_{SUSY}$&  $M_1$  & $M_2$  &  $M_3$  &  $A_t$  & (GeV)\\
\hline
\,\,\,\,\,\,1 & 300  & 100 & 200 & 300 & 500  \\
\,\,\,\,\,\,2 & 500  & 100 & 500 & 800 & 700  \\
\,\,\,\,\,\,3 & 1000  & 100 & 300 & 500 & 1500 \\
\hline \hline
\end{tabular}
\caption{Benchmark points in the NMSSM with $m_a\sim 214$ MeV.}
\label{table1}
\end{table}

\section{Kinematic feature of $a \rightarrow \mu^+\mu^-$}

If we identify $m_a=214.3 \pm 0.5$ MeV, at colliders the behavior of
di-muon as the decay product of $a$ is different from usual cases in
the SM and other physics beyond the SM. In the rest frame of $a$,
di-muon are almost at rest because the mass difference between $m_a$
and $2 m_\mu$ ($2 m_{\mu} \approx$ 211.3 MeV) is tiny, compared to
the typical energy scale of $a$ at colliders. In the lab frame, the
di-muon will be boosted to almost the same direction. As a
consequence, the separation $\Delta R$ of di-muon (see Fig.
\ref{figldr}) is much smaller than usual case. Here $\Delta R$ is
the separation between the two particles in the detector, $\Delta R
\equiv \sqrt{(\Delta \eta)^2 + (\Delta \phi)^2}$; $\phi$ is the
azimuthal angle and $\eta$ denotes pseudo-rapidity.

It should be emphasized that such kind of unusual di-muon events
have always been overlooked in the past experiments. For example,
for the case of the di-muon reconstruction at ATLAS, in order to
suppress the fake muon from pile up and backgrounds etc., one
usually sets cut for $\Delta R$ of the di-muon as 0.01 \cite{BZhou}.
As such the di-muon reconstruction procedure will {\em totally}
abandon the di-muon signal from the decay of $a$. Unfortunately the
LEP \cite{YNGao} and Tevatron experiments may also have the similar
chance to abandon such kind of di-muon. However it is possible
\cite{BZhou,YNGao} that advanced muon detector, especially ATLAS and
CMS, can identify $\Delta R\sim 0$ di-muon with reasonable
efficiency. In fact it is justified to expect that di-muon for
$\Delta R\sim 0$ will have some overlapping hits at first and become
separate tracks in the end, under the strong magnetic field in the
detector. Obviously the efficiency of identifying such kind of
di-muon depends on the details of the whole detector, which is
unknown yet and deserves further detector simulations \footnote{The
very preliminary simulation by Z.C. Yang of CMS collaboration at
Peking University shows that the di-muon tracks are not
distinguishable at pixel detector, but soon become two under the
strong magnetic field. The di-muon reconstruction efficiency is
still substantial, for example $\sim 64\%$ for $p_T(a)=50$ GeV.}.

Provided that the light SM-like Higgs boson $h$ decays dominantly
into $a a$ other than into $b\bar b$ as discussed in section II,
similar to the case of Ref. \cite{Dermisek:2006wr}, and
 $a$ decays dominantly into $\mu^+\mu^-$ in order to account
for HyperCP three events as shown in Ref. \cite{He:2006fr}, the
LEP experiments of direct search for the SM-like Higgs boson can
only impose very weak, even null, constraint on the Higgs boson
mass. The reason is that we have overlooked the signal events due
to the fault of di-muon reconstruction. At LEP, Tevatron and LHC,
one will not discover the SM-like Higgs boson unless the di-muon
reconstruction algorithm and/or the trigger-system are
appropriately realized. Therefore the LEP and Tevatron data should
be re-analyzed. According to the global fit of the precision data
\cite{Chanowitz:2001bv}, the signal of 4$\mu$ ($2\mu^+ 2\mu^-$)
from the decay of the SM-like Higgs boson may lurk in the existing
LEP and/or Tevatron data.

\section{SM-like Higgs boson at colliders}

In this section we will investigate the observability of 4$\mu$
($2\mu^+ 2\mu^-$) as the unique signature of the SM-like Higgs boson
at colliders, especially at Tevatron, LHC and LEP. Throughout the
paper, CP-odd Higgs boson $a$ is assumed to be 0.215 GeV and the
SM-like Higgs boson is the O(100 GeV) or less as implied from
precision data \cite{Chanowitz:2001bv}. The efficiency of
identifying 4$\mu$ from the $h$ decay is taken to be 1. And the real
signal events can be obtained by multiplying with the realistic
efficiency once available.

\subsection{Choice of the parameters}

In our analysis, for simplicity, we assume that the couplings
among the SM-like h and gauge bosons as well as fermions the same
way as those
 in SM. Actually this is the reason why we name the CP-even Higgs boson
'SM-like'. Moreover, as discussed in previous, the new decay mode $h
\rightarrow a a$ needs to be inserted. We assume the h-a-a coupling
as $
\frac{i g m_Z}{2 \cos\theta_W} \kappa $ where $g$, $\theta_W$ are
weak coupling and weak angle as usual and $\kappa$ is dimensionless
free parameter which depends on the specific model. In minimal
supersymmetric standard model (MSSM),
$\kappa=\cos(2\beta)\sin(\alpha+\beta)$ where $\tan\beta$ is the
ratio of two vacuum expectation values of Higgs field and $\alpha$
is the mixing angle of neutral Higgs. In the limit $\sin\alpha
\rightarrow -\cos\beta$ and $\cos\alpha \rightarrow \sin\beta$, the
light CP-even Higgs boson resembles the SM Higgs boson h and $\kappa
\rightarrow -1$ for large $\tan\beta$. Throughout the paper we fix
$\kappa=-1$ in our numerical analysis. In NMSSM the h-a-a coupling
will be altered. However for the light Higgs boson, the different
choice of parameter $\kappa$ in NMSSM affects only the total width
of SM-like Higgs boson  while the branching ratio of $h \rightarrow
aa$ keeps almost the same due to the tiny partial decay width to SM
particles. For relatively heavy Higgs boson which can decay into
$VV^{(*)}$ (V=W or Z) with sizeable branching ratio, the variation
of h-a-a coupling can affect not only the Higgs boson decay width
but also branching ratio of $h \rightarrow aa$. In this paper we
focus on the Higgs boson mass of O(100 GeV) or less and
$Br(h\rightarrow a a) \sim 1$. Therefore the choice of h-a-a
coupling is appropriate for our purpose.

The couplings among $a$ and fermions are taken the same as Ref.
\cite{He:2006fr} in order to account for HyperCP three events,
which can be describe as
\begin{eqnarray}
L_{a \bar f f}= -\frac{i}{v} \left(l_u m_u \bar u \gamma_5 u+l_d
m_d \bar d \gamma_5 d\right) a +  \frac{ig_\ell m_\ell}{v} \bar
\ell \gamma_5 \ell a
\end{eqnarray}
with $ l_d=-g_\ell \sim O(1), l_u=\frac{l_d}{\tan^2\beta} $ in NMSSM
and $v$ the usual VEV of Higgs field. For large $\tan\beta$ case as
preferred by Ref. \cite{He:2006fr}, the up-type quarks have
negligible contributions. In our numerical evaluations, we take
$\tan\beta=30$ and $l_d=-g_\ell=1$, and the branching ratio of decay
$a \rightarrow \mu^+ \mu^-$ is calculated to be $\sim 1$. Note that
$a \rightarrow \pi^0 \gamma$ is forbidden due to C-parity
non-conservation and $a \rightarrow \pi^0 \gamma \gamma$ is assumed
negligible \cite{Hiller:2004ii,SLZHU}. Thus we forbid the decay $a
\rightarrow g g$ in our simulations. In fact branching ratio of
loop-induced process $a \rightarrow \gamma \gamma$ is much less than
that of $a \rightarrow \mu^+ \mu^-$.

\subsection{Detail simulations}

At hadron colliders, namely Tevatron and LHC, the main signal
process for light Higgs boson is $gg \rightarrow h \rightarrow a a
\rightarrow \mu^+\mu^- \mu^+\mu^-$. In our analysis only quark loop
contributions to $gg \rightarrow h$ are included. The signal final
states are $4\mu$ and the main background comes from $ZZ$ production
with Z decaying subsequently into di-muon. In practice, we utilize
Pythia v6.324 \cite{Sjostrand:2006za} to simulate signal and
background after corresponding modifications of couplings and
masses.

In our simulations, we adopt the following ``LHC basic cuts'' from
\cite{Zhu:2005hv}:
\begin{equation}
    p_T(\mu^{\pm}) > 10 \ {\rm GeV}, \
    |\eta(\mu^\pm)| < 2.5. \
\label{lhc1}
\end{equation}
 At Tevatron we adopt the
following ``Tevatron basic cuts'' from \cite{Zhu:2005hv}:
\begin{eqnarray}
   & p_T(\mu^{\pm}) > 12 \ {\rm GeV}, \
    |\eta(\mu^\pm)| < 2.0. \
\label{tev1}
\end{eqnarray}

\begin{figure}[thb]
\vbox{\kern2.5in\includegraphics{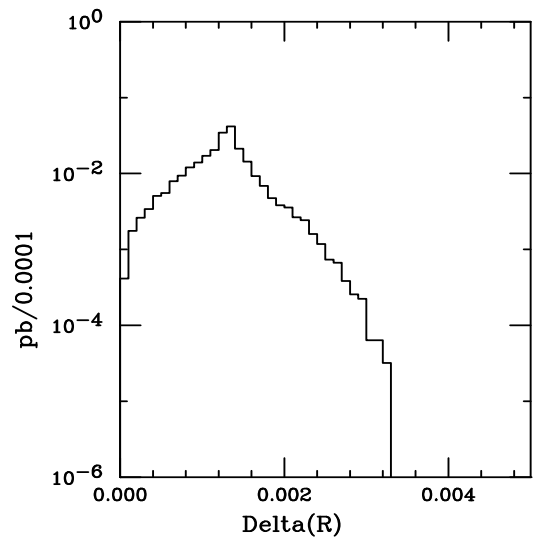}} \caption{Distribution of $\Delta
R$ between $\mu^+\mu^-$ at Tevatron for signal $gg \rightarrow h
\rightarrow a a \rightarrow 4 \mu$ with "Tevatron basic cuts" of Eq.
(\ref{tev1}). In all figures of this paper, $m_h=120$ GeV and
$m_a=0.215$ GeV, except indicated otherwise. }\label{figldr}
\end{figure}
In order to demonstrate the unique kinematics of di-muon, in Fig.
\ref{figldr} we show the separation $\Delta R$ between
$\mu^+\mu^-$ from $a$ decay at Tevatron with $\sqrt{s}=2$ TeV. It
should be noted that the result at LHC is similar to that at
Tevatron. From the figure we can see clearly that $\Delta R$ is
much less than the usual ATLAS cut (say 0.01 \cite{BZhou}), in
order to suppress the fake muons. Such kind of di-muon has not
been noticed, at least not emphasized, in the past and/or on-going
analysis. In order to keep the unique $4\mu$ signal, special
attention to the extremely small $\Delta R$ of di-muon should be
paid.


\begin{figure}[thb]
\vbox{\kern2.6in\includegraphics{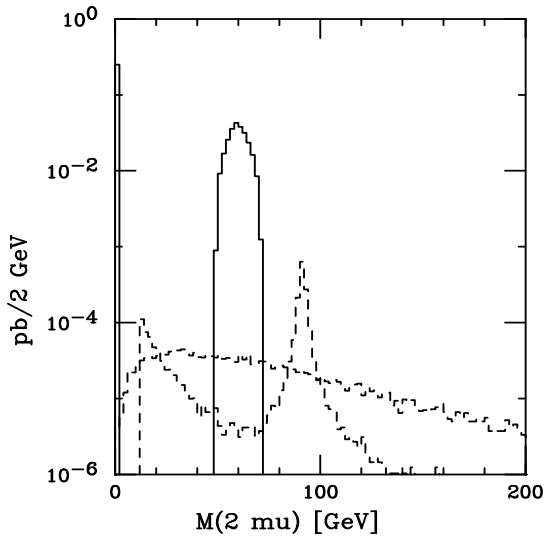}} \caption{Distributions of
invariant mass for di-muon at Tevatron for signal $gg \rightarrow h
\rightarrow a a \rightarrow 4 \mu$ and for background $q \bar q
\rightarrow ZZ \rightarrow 4 \mu$ with "Tevatron basic cuts".The
solid (dashed) lines in all figures represent signal (background).
}\label{2mu}
\end{figure}
Choosing any one $\mu^+$ from the two, we can have two $\mu^+\mu^-$
combinations. In Fig. \ref{2mu} the invariant mass $M_{\mu^+\mu^-}$
of signal and background for two combinations at Tevatron are shown.
The results at LHC are similar. For signal, the enhancement around
$m_a$ will be smeared due to the limited detector energy resolution
and the very tiny decay width, which is O($10^{-7}$) MeV. The other
peak is around $m_h/2$ because of the characteristic of the nearly
collinear $\mu^+\mu^-$ from the same $a$ decay, i.e. $
M_{\mu^+\mu^-}\simeq \frac{1}{2} M_{aa} \simeq \frac{1}{2} m_h. $
Clearly $M_{\mu^+\mu^-}$ arising from the background for which
di-muon comes from Z has peak around Z-mass, and $M_{\mu^+\mu^-}$
for another combination is continuous distributed. Such properties
can be utilized to suppress background. Namely we can exclude
di-muon invariant mass around $m_Z$. However this cut will bring
potential risk if $m_h$ is around $2 m_Z$.

\begin{figure}[thb] \vbox{\kern2.6in\includegraphics{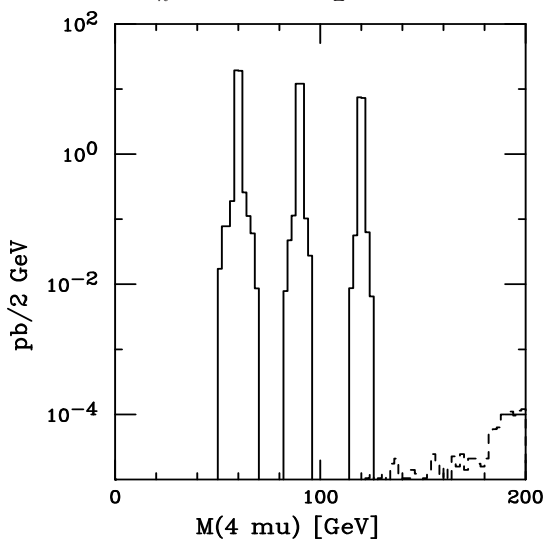}} \caption{ Distributions of invariant mass of four
$\mu$ for signal and  background at LHC with $\sqrt{s}=14$ TeV.
"LHC basic cuts" are applied. Here the SM-like Higgs boson mass is
taken to be 60, 90 and 120 GeV respectively.}\label{invlhc}
\end{figure}
\begin{figure}[thb]
\vbox{\kern2.6in\includegraphics{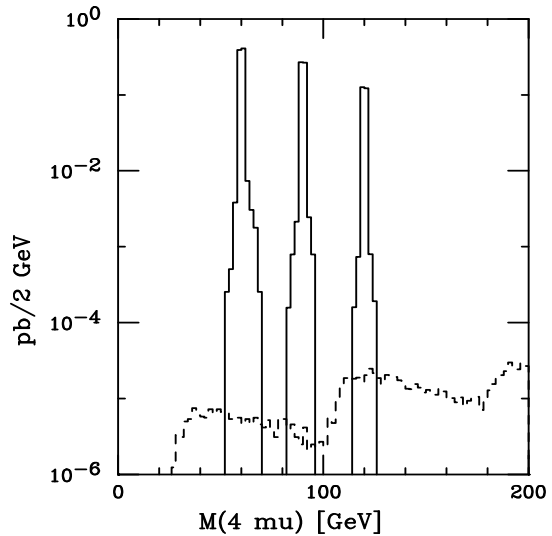}} \caption{Same with Fig.
\ref{invlhc} but at Tevatron with "Tevatron basic cuts".
}\label{invtev}
\end{figure}
In Fig. \ref{invlhc} and Fig. \ref{invtev}, we present the 4$\mu$
invariant mass distribution for signal and background at LHC and
Tevatron respectively. Here $m_h$ is taken to be 60, 90 and 120
GeV. It is clear that SM-like Higgs boson mass can be precisely
reconstructed and 4$\mu$ background is rather low. This conclusion
won't change provided that the efficiency of 4$\mu$ reconstruction
is not extremely low. At Tevatron with 1 $fb^{-1}$ integrated
luminosity, for $m_h=120$ GeV, we will have 250  signal events.
Even the real efficiency for $4\mu$ reconstruction is 10\%, we
still have 25 events. At LHC for the same luminosity, Higgs boson
mass and reconstruction efficiency, we can have $\sim 1500$ signal
events. Provided the rather low $4\mu$ background from SM, it is
very promising to discovery the Higgs boson at LHC and Tevatron
via the unique $4\mu$ mode with rather low luminosity.

\begin{figure}[thb] \vbox{\kern2.6in\includegraphics{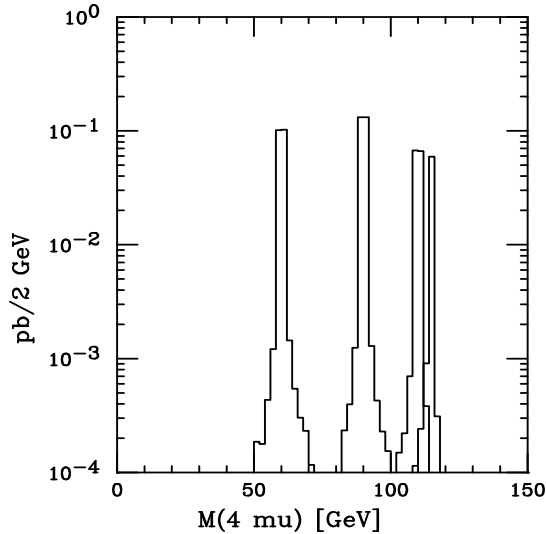}} \caption{ Distributions of invariant mass of four
$\mu$ for $e^+ e^- \rightarrow Z h \rightarrow Z+ 4\mu$ at LEP with
$\sqrt{s}=208$ GeV. Requirements of Eq.(\ref{tev1}) are applied.
Here the SM-like Higgs boson mass is taken to be 60, 90 110 and
115 GeV respectively.}\label{invlep}
\end{figure}
At linear colliders  h production  is mainly via $e^+ e^-
\rightarrow Z h$. The SM-like Higgs boson h then subsequently
decays into $aa$ and $aa \rightarrow \mu^+\mu^- \mu^+\mu^-$, and Z decays into
anything other than $\mu^+\mu^-$ in order not to mix with the $4\mu$ from h. The
background for this process is also rather low. Therefore in Fig.
\ref{invlep} we show only invariant mass of $4\mu$ from signal.
For $\sqrt{s}=208$ GeV, with 500 $pb^{-1}$ luminosity and 10\%
efficiency for $4\mu$ reconstruction, we will have about 3, 7, 13
and 10 events for $m_h=115, 110,90$ and 60 GeV respectively.

\section{Discussions and conclusions}

 To summarize, we investigate in this paper the unique 4$\mu$
signature of the SM-like Higgs boson h at colliders, i.e. $h
\rightarrow a a \rightarrow \mu^+\mu^- \mu^+\mu^-$. Here $a$ is
supposed to be the narrow-width CP-odd scalar with the mass $214.3
\pm 0.5$ MeV, which can account for HyperCP three di-muon events. As
the consequence of tiny mass difference between $m_a$ and 2$\mu$,
di-muon will be boosted to almost the same direction at colliders.
Such kind of di-muon events have been overlooked in the past
experiments. Provided that the SM-like Higgs boson $h$ decays
dominantly into $a a$ other than into $b\bar b$, in order to be
consistent with null Higgs boson search at LEP, the $h\rightarrow
 4\mu$ will be the unique Higgs boson signature
which has not been investigated before. The SM-like Higgs boson
may hide itself from the usual analysis of LEP and Tevatron
experiments, which should be reanalyzed in the light of new
theoretical and experimental developments. In this paper, we
investigate this unique Higgs boson signature at colliders in, but
actually not limited to, NMSSM and conclude that the SM-like Higgs
boson could be discovered with rather low integrated luminosity,
provided that the $h\rightarrow 4\mu$ reconstruction efficiency is
not extremely low. It is not impossible that such kind of unique
Higgs boson $4\mu$ events are now lurking in the existing LEP
and/or Tevatron data. Moreover we are aware that the
simulation presented here is very rough and the full detector
simulation is the natural further investigation.

 {\em Acknowledgements}: The author thanks  Yong
Ban, Yuan-ning Gao and Bing Zhou for the useful communications on
muon reconstruction, Shi-lin Zhu for the discussions on $a$ decays
into hadronic states, Chong-shou Gao and Chuan Liu for the
stimulating discussions on possible theoretical interpretations for
HyperCP three events in 2005 summer.
 This work was supported in part by the Natural Sciences Foundation of
China (No. 90403004 and 10635030), the trans-century fund and the
key grant project (No. 305001) of Chinese Ministry of Education.

\end{document}